# An enhanced merger fraction within the galaxy population of the SSA22 protocluster at $z = 3.1$

N. K. Hine[⋆1], J. E. Geach[1], D. M. Alexander[2], B. D. Lehmer[3,4,5], S. C. Chapman[6], Y. Matsuda[7,8]

[1]*Centre for Astrophysics Research, Science and Technology Research Institute, University of Hertfordshire, Hatfield AL10 9AB, UK*
[2]*Centre for Extragalactic Astronomy, Department of Physics, Durham University, South Road, Durham DH1 3LE, UK*
[3]*Department of Physics and Astronomy, The John Hopkins University, Baltimore, MD 21218, USA*
[4]*NASA Goddard Space Flight Center, Code 662, Greenbelt, MD 20771, USA*
[5]*University of Arkansas, 226 Physics Building, Fayetteville, AR 72701, USA*
[6]*Department of Physics and Atmospheric Science, Dalhousie University, Halifax, NS, B3H 3J5, Canada*
[7]*National Astronomical Observatory of Japan, 2-21-1 Osawa, Mitaka, Tokyo 181-8588, Japan*
[8]*The Graduate University for Advanced Studies (SOKENDAI), 2-21-1 Osawa, Mitaka, Tokyo 181-0015, Japan*

**ABSTRACT**
The overdense environments of protoclusters of galaxies in the early Universe ($z > 2$) are expected to accelerate the evolution of galaxies, with an increased rate of stellar mass assembly and black hole accretion compared to co-eval galaxies in the average density 'field'. These galaxies are destined to form the passive population of massive galaxies that dominate the cores of rich clusters today. While signatures of the accelerated growth of galaxies in the SSA22 protocluster ($z = 3.1$) have been observed, the mechanism driving this remains unclear. Here we show an enhanced rate of galaxy-galaxy mergers could be responsible. We morphologically classify Lyman-break Galaxies (LBGs) in the SSA22 protocluster and compare these to those of galaxies in the field at $z = 3.1$ as either active mergers or non-merging using *Hubble Space Telescope* ACS/*F*814*W* imaging, probing the rest-frame ultraviolet stellar light. We measure a merger fraction of 48±10 per cent for LBGs in the protocluster compared to 30±6 per cent in the field. Although the excess is marginal, an enhanced rate of mergers in SSA22 hints that galaxy-galaxy mergers are one of the key channels driving accelerated star formation and AGN growth in protocluster environments.

## 1 INTRODUCTION

Most of the stars in the Universe were formed during a peak star formation era at $1 \leqslant z \leqslant 3$ (Madau et al. 1996; Sobral et al. 2013) when the volume averaged star formation rate (SFR) density was about 10 times higher than today. The strong evolution in the average rate of galaxy growth is thought to be driven by a combination of higher merger rates (Somerville et al. 2001; Conselice et al. 2003; Hopkins et al. 2006; Conselice 2014) and a higher rate of gas accretion on to dark matter haloes (Keres et al. 2005; Dekel et al. 2009) resulting in large, turbulent discs with high gas fractions compared to today (Geach et al. 2011; Swinbank et al. 2011; Genzel et al. 2013; Tacconi et al. 2013). Understanding the relative importance of the different processes driving galaxy growth and evolution, and the balance between them as a function of local environment, is a key focus of galaxy evolution studies.

Either directly or indirectly, environment is known to be a major influence on the evolution of galaxies. In the local Universe, the most massive galaxies are located at peaks in the density field. The progenitors of $z \sim 0$ clusters are overdense regions at higher redshift, although not necessarily virialized or dominated by quiescent populations (Steidel et al. 1998). Such protoclusters have been detected around massive high redshift radio galaxies at z⩾2 (Le Fevre et al. 1996; Venemans et al. 2002, 2004, 2005, 2007; Stern et al. 2003; Hatch et al. 2009; Matsuda et al. 2009). However, protoclusters have also been identified 'blindly' as significant overdensities of Lyman break galaxies (LBGs) and Ly$\alpha$ emitters (LAEs) in regions where no massive radio galaxy is detected (Steidel et al. 1998, 2000; Ouchi et al. 2005; Matsuda et al. 2010). Protoclusters could have formed due to the preferential accretion of gas collapsing on to dark matter filaments and nodes, leading to the more rapid formation of stars and galaxies compared to average density regions at a similar redshift (Matsuda et al. 2005). It is predicted that galaxies will evolve more quickly in these regions through a combination of accelerated infall of gas and a higher rate of mergers (Kauffmann 1996; De Lucia et al. 2006; Gottlöber et al. 2001; Fakhouri & Ma 2009). At a given epoch, we might therefore expect to see galaxies in protoclusters at a later stage of evolution, or in a more rapid phase of growth than galaxies in the field (Lehmer et al. 2009; Steidel et al. 2005; Kubo et al. 2013). The densest protoclusters at high-$z$ are expected to evolve into the most massive clusters seen in the local Universe (Governato et al. 1998; De Lucia et al. 2006). Studying galaxies in protocluster environments therefore allows us to explore the early history of the most massive early-type galaxies today, as well as testing the hypothesis that evolution is accelerated in dense environments at early times. For example, Lotz et al. (2013) found an enhanced merger fraction in a protocluster at z = 1.62 and we test for a similar enhancement at higher redshift.

⋆ E-mail:n.hine@herts.ac.uk

Galaxy mergers increase galaxy mass and trigger starbursts through the collapse of molecular clouds. As angular momentum is dissipated, gas is also channeled into galactic nuclei where it can fuel supermassive black holes radiating as AGN. In this paper, we compare the merger fraction in the SSA22 protocluster at $z\sim3$ with that in *Hubble* Deep Field North (HDF-N), considered to be a region of average density. We test the hypothesis that the merger fraction in protocluster environments is higher than compared to the average density field and therefore potentially responsible for accelerated growth in these environments. In Section 2 we introduce the SSA22 protocluster and describe samples and source data. Section 3 sets out our approach to classifying the galaxies as merging or non-merging and gives the results of our work. In Section 4 we discuss the interpretations and limitations before presenting our conclusions in §5. All magnitudes are on the AB system and a $\Lambda$ cold dark matter ($\Lambda$CDM) cosmology of $\Omega_m = 0.3$, $\Omega_\lambda = 0.7$ and $H_0 = 70$ km s$^{-1}$ Mpc$^{-1}$ is assumed throughout.

## 2 DATASETS

### 2.1 The SSA22 protocluster

The SSA22 protocluster (R.A. = 22h 17m, Dec. = +00° 15′) was first discovered by Steidel et al. (1998) as a spike in the redshift distribution of LBGs at $z = 3.1$, since found to be six times as dense as the field at this redshift (Steidel et al. 2000). Later Narrow band Ly$\alpha$ imaging has identified additional $z = 3.1$ galaxies (LAEs) in a larger region around the original spike, extending for more than 60 Mpc comoving (Hayashino et al. 2004; Yamada et al. 2012) and thought to trace several dark matter filaments intersecting to form a density peak (Matsuda et al. 2005). Overdensities of Ly$\alpha$ absorbers (Hayashino et al. 2004), Ly$\alpha$ blobs (Steidel et al. 2000; Matsuda et al. 2004, 2011), AGN (Lehmer et al. 2009) and submillimeter galaxies (Geach et al. 2005, 2014; Tamura et al. 2009; Umehata et al. 2014) have also been detected.

SSA22 is thought to be a rare high-density peak that will evolve to form a massive ($10^{15}$M$_\odot$) cluster by $z = 0$. Recent studies (Kubo et al. 2013, 2015) have found that a significant proportion of massive galaxies in the protocluster have already become quiescent, although star formation and AGN activity still dominate, implying that the build up of the massive galaxy population is still ongoing. This indicates that the protocluster represents a key stage in the evolution of massive galaxies as they evolve from star forming to quiescence. Lehmer et al. (2009) found evidence that the LBGs in SSA22 are $\geqslant 1.2 - 1.8$ times as massive as those in the field and contain a higher fraction of AGN compared to the field at the same epoch. These observations suggests that galaxy growth is accelerated in this dense region. As such it provides an opportunity to study the mechanism of the acceleration.

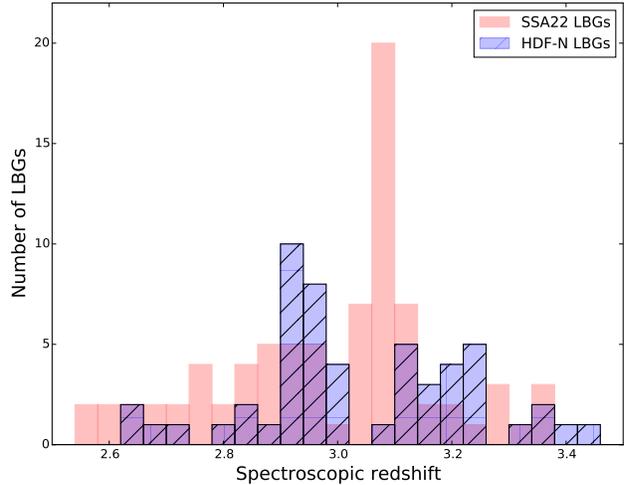

**Figure 1.** The redshift distribution of LBGs in SSA22 and HDF-N fields, the protocluster lies at $z = 3.09$. Redshifts were obtained from the Steidel et al. (2003) catalogues.

### 2.2 Lyman Break Galaxy sample selection

Our sample is taken from the Steidel et al. (2003) LBG redshift survey using the Palomar 5.08-m telescope. The LBGs span a magnitude range of $19.0 \leqslant R_{AB} \leqslant 25.5$ mag, and spectroscopic redshifts have been obtained using the Low Resolution Imaging Spectrometer (LRIS), on Keck (Oke et al. 1995). We identify LBGs at $z = 3.06 - 3.12$ as members of the protocluster (Lehmer et al. 2009); as this is not a virialized structure, the concept of 'membership' is not well defined. For the average density field control sample we select LBGs in the HDF-N field in the range $2.5 \leqslant z \leqslant 3.5$. To expand our control sample we also classified the SSA22 LBGs at $2.5 \leqslant z \leqslant 3.5$, but excluding the $3.06 \leqslant z \leqslant 3.12$ interval, as 'field' galaxies not associated with the protocluster. The redshift distribution of both samples is shown in Fig. 1.

We tested the similarity of the two LBG samples using IRAC 4.5$\mu$m magnitudes ([4.5]) as a proxy for stellar mass. At $z = 3.1$ MIR bands trace the lower mass stars which make up most of the baryonic mass in galaxies. Where available we obtained IRAC data for the HDF-N sample from Ashby et al. (2013) and for the SSA22 sample from the *Spitzer* archive (originally from GO project 30328). Applying the Kolmogorov–Smirnov (K-S) test gave a p-value of 0.97 for [4.5] < 24, (where both catalogues are approximately complete), confirming that the two samples have a statistically identical stellar mass distribution.

### 2.3 *HST* observations

We obtained archival *Hubble Space Telescope* (*HST*) images for the LBGs using the Mikulski Archive for Space Telescopes (MAST)[1]. All images were acquired using the Advanced Camera for Surveys (ACS) and the *F*814*W* filter. We were able to obtain images for

---

[1] http://archive.stsci.edu/. Proposal numbers 9760, 10405, 11636, 12527, 12442, 12443, 12444, 12445, 13063 and 13420.

23 of the 27 LBGs in the protocluster and 33 of the 55 LBGs in the SSA22 field sample given the sparse sampling strategy of the *HST* projects. There was full coverage of the HDF-N field LBGs. Exposure times for individual protocluster images varied from 2–7.3ks (with some combined to give total exposures of up to 11ks) and exposure times for the field samples were restricted to a similar range. The potential impact of the range of exposure times used is discussed further in Section 3.4.

# 3 ANALYSIS

## 3.1 Classification

We generated 4 arcsec × 4 arcsec (≈ 30 kpc projected) thumbnail images centred on the coordinates of each LBG where image data of suitable quality was available (e.g. not on an edge). For the classification, we scale each of these images to produce three versions, containing (a) all flux between ∼1 and 15 per cent of the peak flux value (faint scale); (b) all flux between ∼15 and 50 per cent of the peak flux value (medium scale) and (c) all flux between ∼55 and 80 per cent of the peak flux value (bright scale). We selected these flux ranges to facilitate the visual identification of different structures associated with mergers, such as faint tidal features, coalescing nodes, and so on.

We used the scaled thumbnails for our initial classification, but we also examined the full dynamic range of each image to aid our classification. All targets were classified at least three times, including once by an independent reviewer (see Section 3.3), with additional reviews carried out for those where the classification was ambiguous. We also carried out a 'blind' test (see Section 3.3) to eliminate potential unconscious bias. We define six classification categories:

C1  Compact, isolated system, single nucleus.
C2  Compact but with minor asymmetry and no clear evidence of a second nucleus.
M1  Evidence of two nuclei within 1 arcsec diameter aperture (∼8kpc in projection) centred on the brightest part of the LBG. All flux falling within the aperture.
M2  Evidence of two nuclei within the 1 arcsec aperture, but with some flux falling outside.
M3  Evidence of >2 nuclei or complex clumpy structure falling within the aperture, but with some flux falling outside.
M4  Evidence of >2 nuclei or complex clumpy structure with all flux falling inside the 1 arcsec diameter aperture.

Those classified as 'C' are non mergers, whilst those classified as 'M' are mergers. A small number of targets could have been classified as C2 or a merger. Where there was doubt we conservatively classified as C2, requiring a clear secondary concentration of bright flux (thought to be a nucleus) to classify as a merger. A small number of thumbnails included a near neighbour outside the aperture which could possibly have been interacting with the target LBG in an early stage merger. We did not classify these as a merger unless there was evidence of diffuse emission reminiscent of stellar tidal trails between the two galaxies. The observed morphology could also have been effected by the impact of dust on the UV emission, this is discussed further in Section 4. Thumbnails of the protocluster LBGs and their classifications are presented in Fig. 2. The images include contours representing the 15, 55 and 80 per cent scaling boundaries used in our analysis, which give a good indication of our classification in most cases.

## 3.2 Merger fractions and merger rates

Merger rates, as defined in the literature, are difficult to calculate; they require time-scales which are generally obtained from simulations (although see Conselice 2009). For example Lotz et al. (2011) defined the merger rate as:

$$\Gamma_{\rm m} = \frac{f_{\rm gm} \times n_{\rm gal}}{T_{\rm obs}} \quad (1)$$

where $f_{\rm gm}$ is the galaxy merger fraction, $n_{\rm gal}$ is the comoving number density of galaxies and $T_{\rm obs}$ is the average observability time-scale. However, *merger fractions* are relatively easy to measure as they are a simple expression of the instantaneous number of mergers observed in a population and so do not require a time-scale. Two merger fractions are used in the literature (Conselice et al. 2008; Lotz et al. 2011; Stott et al. 2013), the most common and the one used in this work is simply:

$$f_{\rm m} = \frac{N_{\rm m}}{N_{\rm t}} \quad (2)$$

where $N_{\rm m}$ is the number of observed mergers in a sample and $N_{\rm t}$ is the total number of galaxies. The second is $f_{\rm gm}$ (the galaxy merger fraction) which is based on the number of galaxies undergoing mergers rather than the number of merging events. In the simple case of two galaxies merging $f_{\rm gm} = 2 f_{\rm m}$.

Comparing merger fractions from different studies is problematic as they are highly dependent on the method used to identify mergers, which is often subjective. In our work we use the same classification method for the protocluster and field samples and therefore the fractions should be comparable across the different environments. We calculate merger fractions based on our classification scheme described above. It is important to note that our scheme detects mergers at a specific phase of the merger sequence in all fields, when the two nuclei are very close, but have not yet coalesced.

Our classification results are presented in Table 1 and Figs 4 and 3. We measure a merger fraction of 48±10 per cent for the protocluster, 33±8 per cent for the SSA22 field and 30±6 per cent for the HDF-N field, assuming binomial statistics to estimate the uncertainties (Berendsen 2011). Combining the two field samples by simply adding the number of galaxies classified as mergers or non-mergers gives a result of 31±5 per cent. Based on these statistics we estimate a probability of ∼7 per cent for finding 11 or more mergers in the protocluster sample if the actual merger fraction is 31 per cent (as in the combined field).

## 3.3 Blind and Independent Testing

To check for unconscious bias in our classification process we generated random IDs for all our scaled images and reclassified them without knowing their provenance (i.e. protocluster or field). 95 per cent of our classifications were unchanged under blind classification. The classification changed for five galaxies: three changing from mergers to non mergers and two from non-mergers to mergers. Due to the small number of galaxies in the protocluster sample, the two protocluster galaxies that changed from mergers to non-mergers do have a small but not significant impact on the merger fraction, which becomes 39±10 per cent. However, these two galaxies were also classified as non-mergers during our initial review of the scaled images and it was only during our more detailed examination that we identified the features that led to their final classification as mergers. The blind testing does not identify any systematic bias in our classifications.

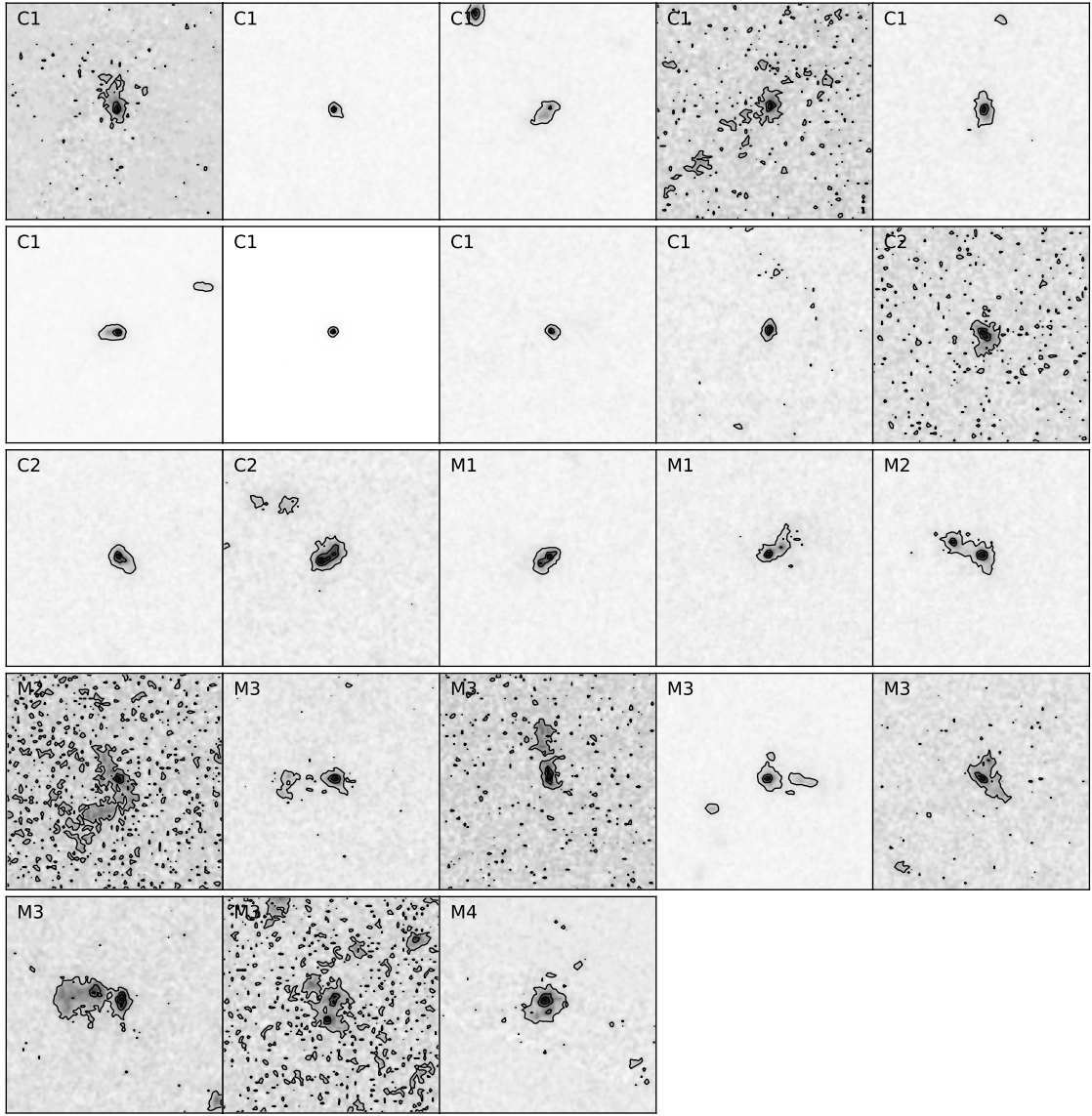

**Figure 2.** 4 arcsec × 4 arcsec (approximately 30 kpc projected) thumbnail ACS *F*814*W* images of the protocluster LBGs grouped and labelled by classification. The contours are at the 15, 55 and 80 per cent levels used in our analysis.

An additional independent merger analysis was carried out by a student who was not a member of the group. The independent analysis indicated a protocluster merger fraction of 43±10 per cent and a combined field merger fraction of 22±4 per cent. Both values are lower than the original results, but the independent analysis still indicates a marginally higher merger fraction in the protocluster, with only a 2 per cent probability of obtaining these results if the merger fractions were actually the same.

### 3.4 Chance alignments and image depth

It is possible that when a second source falls within the 0.5 arcsec radius aperture used to categorise the LBGs, this source is not an interacting galaxy, but simply a chance alignment of physically disassociated galaxies along the line of sight. We investigated this possibiliy using the following procedure. First, we evaluated the median LBG flux within a 0.5 arcsec radius aperture. We then randomly selected $10^4$ pixels in each of the full frame images, each time evaluating the total flux within 0.5 arcsec of that pixel. This procedure allows us to estimate the surface density of objects of similar (or greater) flux to the target LBGs, and thus a means of evaluating the likelihood of chance alignments. The results show that for each LBG, there is only ∼1 per cent probability that an object of similar flux would be detected within 0.5 arcsec (i.e. mimicking a merger).

Our analysis involved the use of a range of images with different exposure times. It is possible that this could also lead to bias in our results, in particular in identifying faint tidal emission that might only become apparent in deeper images. For the SSA22 sample the exposure times of images used range from 2.2–11 ks, and for HDF-N from 1.2–8.9 ks. There were no mergers identified in the $\sim$2ks images, but as there were only two thumbnails in this group, this is unlikely to be significant. The merger fraction for all LBGs in the SSA22 field and protocluster combined was 39 per cent, but when split by exposure times the merger fractions varied between 44 per cent and 33 per cent (ignoring the $\sim$2ks group) showing a slight fall with increasing exposure time. The HDF-N images show a stronger bias, but in the opposite direction. We measured a merger fraction of 26 per cent for shorter exposures ($<$ 3ks) and 39 per cent for longer exposures ($>$ 8ks).

**Table 1.** Morphological classification and merger fractions for LBGs selected in the protocluster and field environments. Uncertainties on the merger fraction are derived from binomial statistics.

| Field | C1 | C2 | Total non-mergers | M1 | M2 | M3 | M4 | Total mergers | Sample size total | Merger fraction |
|---|---|---|---|---|---|---|---|---|---|---|
| SSA22 Protocluster | 9 | 3 | 12 | 2 | 2 | 6 | 1 | 11 | 23 | 0.48±0.10 |
| SSA22 Field | 17 | 5 | 22 | 3 | 2 | 5 | 1 | 11 | 33 | 0.33±0.08 |
| HDF-N Field | 26 | 11 | 37 | 4 | 7 | 5 | 0 | 16 | 53 | 0.30±0.06 |
| Combined field | 43 | 16 | 59 | 7 | 9 | 10 | 1 | 27 | 86 | 0.31±0.05 |

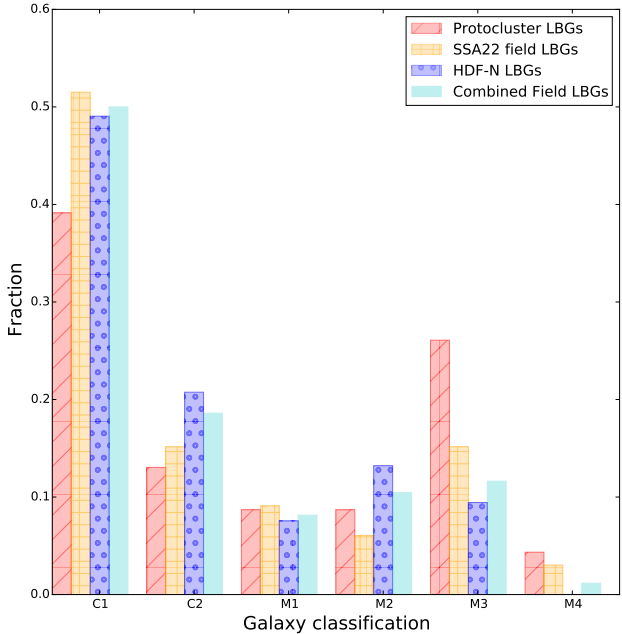

**Figure 3.** Fraction of galaxies of each type for the protocluster and average field samples.

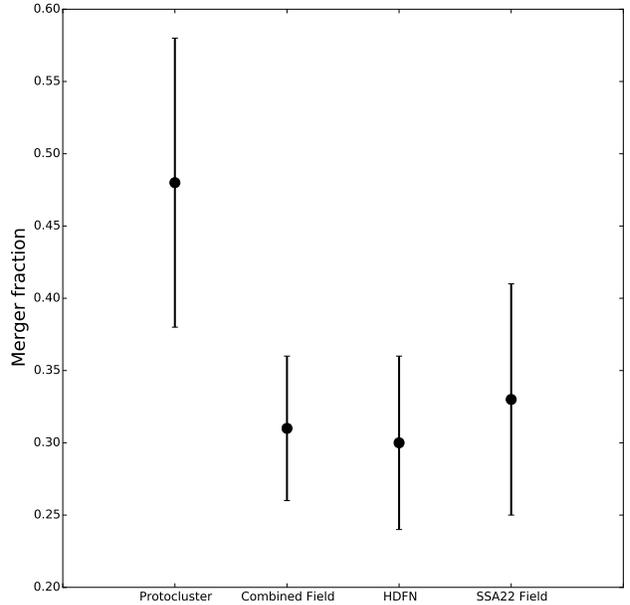

**Figure 4.** Merger fractions for our individual samples and the combined field population based on observed morphology.

We also compared the percentage of mergers in each exposure time band with the percentage of thumbnails using that exposure time. This indicates a slight enhancement in the SSA22 merger fraction in long exposures, but it is not significant enough to affect our conclusions. For HDF-N we found that 34 per cent of the mergers were in the longer exposures which make up 44 per cent of all thumbnails, giving lower than expected occurrence in long exposures. Thus there was no consistent bias.

All the mergers initially identified using long-exposure images would also have been identified as mergers if we had used the available short-exposure images. All but two of the compact classifications identified in the long-exposures would also have been identified as compacts using short exposures, and two might have been misidentified as mergers due to noise that appeared to link two separate objects. These tests give us confidence that our classification scheme is not strongly dependent on the depth of the image.

### 3.5 Comparison to CAS and Close Pairs

We measured 'CAS' (concentration, asymmetry, smoothness/clumpiness) parameters (Conselice 2008) for our LBGs with a view to determining CAS based merger fractions for comparison. However, these were not able to distinguish between the merging and non merging systems in our sample (C. Conselice, private communication). The $A$ parameter is considered to be the most effective for measuring mergers ($A > 3.5$ indicates a merger, Conselice et al. 2008). This definition gives a merger fraction of 22 per cent for the protocluster and 4 per cent for the HDF-N field (the distribution of $A$ values is shown in Fig. 5). This does indicate an enhanced merger fraction in the protocluster, but is clearly only identifying a small number of the mergers in our classification (seven in total). This is perhaps not surprising given the limited resolution and relatively low signal to noise of the images at $z \approx 3$ (a signal to noise of 20 is required, C. Conselice, private communication) as well as the fact that our definition of a merger is based on the presence of two unmerged nuclei rather than disturbed morphology.

We carried out a close pairs analysis, using a separation of 20kpc. This resulted in a merger fraction of 52±10 per cent. in the protocluster and 38±5 per cent in the combined field. The larger aperture leads to an increased probability of chance alignment of ∼3 per cent. The probability of the merger fraction in the protocluster being the same as that in the field based on these results increases to 12 per cent. Overall we consider these results to be less reliable due to the increased chance of a low redshift interloper.

Studies of massive galaxies at slightly lower redshift ranges (2.3 < z < 3.0) have found field merger fractions based on close pair analysis (at 30kpc) of 40±10 per cent (Bluck et al. 2009) and 12±15 per cent (Man et al. 2012). The large uncertainties mean that these findings are consistent with each other and with our results for our field samples. A similar study using CAS parameters indicated a field merger fraction of 27±8 per cent (Bluck et al. 2012), which is also consistent with our results and with earlier work suggesting a peak in the field merger fraction of ∼30 per cent at $z$=3 (Conselice & Arnold 2009).

These studies involved rest-frame optical observations, whereas our data are rest-frame UV. Taylor-Mager et al. (2007) investigated the impact of using different rest-frame wavelengths to classify galaxies using the CAS parameters and found that using UV data led to higher values for A. Taylor-Mager et al. (2007) found that nearby late-type galaxies (including mergers) were less concentrated, more asymmetric, and more clumpy when observed at UV wavelengths compared to optical observations. They suggest that an adjustment needs to be made to the CAS parameters when applying them to UV observations. Conselice et al. (2008) also found a fractional change in the A parameters when observed in UV rather than optical light of $\frac{\Delta A}{\Delta \lambda} = -0.83 \pm 1.06$ at $z = 0.75 - 1.25$. Given the low number of mergers identified in Section 3.5 it seems unlikely that the $A$ parameter is identifying too many mergers as this lower redshift work would predict. Huertas-Company et al. (2014) found there was generally good correlation between morphological type for UV and optical observations up to $z = 3$. They used a Support Vector Machine with seven parameters (including CAS) to determine the probability of a particular galaxy type and found a relationship of $p^{uv}_{\rm irregular} = (0.7 \pm 0.02) \times p^{\rm optical}_{\rm irregular} + (0.13 \pm 0.01)$. This suggests that wavelength may not have as large an impact as had previously been thought.

## 4 INTERPRETATION AND DISCUSSION

Our classification method is designed to be sensitive to *ongoing* mergers at a specific stage in the merger sequence; very late stage mergers (close to coalescence) or very early (widely separated) interactions are counted as 'non-mergers'. The experiment was intended to be conservative such that we only distinguish between galaxies that are clearly comprised of two or more compact nuclei separated by several kiloparsecs, or isolated systems. Importantly, however, the same methodology was applied to the protocluster members and the field. Our results indicate a marginal enhancement of the number of on-going mergers in the protocluster environment at $z \approx 3$. Aside from the possibility that this enhancement merely reflects a statistical fluctuation (described in §3) there are three possible physical explanations for this:

(i) *Higher merger rate.* This implies that either more galaxies are undergoing mergers at any one time in the protocluster compared to the field, or that there is a mis-match between the formation epoch of the protocluster and field LBGs.

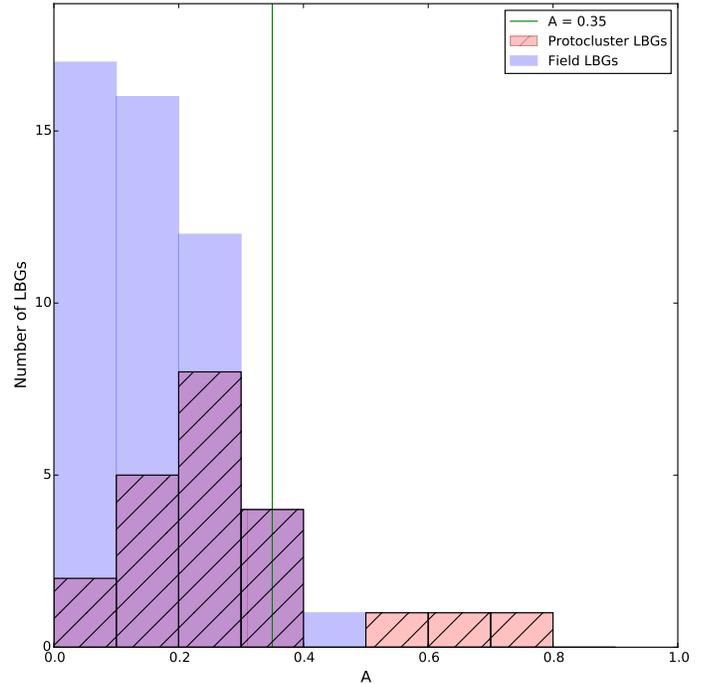

**Figure 5.** The $A$ parameter values for the SSA22 protocluster and HDF-N field LBGs. The vertical line indicates the Conselice et al. (2008) classification of a merger. Using this classification only 7 LBGs in total were identified as mergers.

(ii) *Multiple mergers.* A higher merger rate could also be due to individual galaxies undergoing multiple mergers, rather than more galaxies undergoing mergers at the same time. We have assumed that each LBG undergoes only one major merger.

(iii) *A longer duty cycle.* Integrated over time, the merger rate may be the same in the protocluster as in the field, but the protocluster galaxies may take longer to merge compared to their field counterparts, increasing their visibility.

The duty cycle of a merger is thought to be dependent on local factors such as galaxy mass (affecting the dynamics of the merger) and gas fraction (Hopkins et al. 2006; Lotz et al. 2008). Under normal conditions, the rate of star formation scales with gas fraction, so provided dust-obscuration is not extreme, gas-rich galaxies undergoing mergers might be detectable for longer (e.g. through stellar streams, bright cores) than gas-poor galaxies. CO surveys of protocluster and identically-selected field galaxies would allow one to compare the gas mass fractions of these two populations. Protocluster galaxies might be expected to have higher gas fractions due to efficient supply of pristine intergalactic gas through filamentary accretion, but this is a picture yet to be empirically tested. Ly$\alpha$ blobs are generally found in high-$z$ protoclusters and indicate the presence of large volumes of cold gas around forming galaxies.

Fakhouri & Ma (2009) examined the role of environment on the halo merger rate in *N*-body simulations, using the friend-of-friends group finder (Davis et al. 1985) within the millennium simulation (Springel et al. 2005) to trace the merger histories of dark matter sub-haloes out to $z = 2$. A key finding was a correlation between environment and merger rate, such that sub-haloes in the densest environments underwent merger rates 2–2.5 times higher than those in the lowest density regions. With the caveat that galax-ies are biased tracers of the matter field, it is reasonable to assume that there will be a similar trend for the galaxies themselves. It is interesting that the level of enhancement of the merger rate found by Fakhouri & Ma (2009) in the simulations is broadly in agreement with the enhancement we find for SSA22.

Our analysis traces the rest-frame ultraviolet emission which is very sensitive to dust obscuration; for example, extended/clumpy star-forming regions could be hidden and the merger fraction understated, or a clumpy dust distribution could lead to very high extinction to certain stellar populations, giving the stellar emission overall a clumpy morphology, that could possibly be misinterpreted as a merger.

We do not expect the use of rest-frame UV observations to lead to a significant bias in our interpretation unless there is a significant difference between the obscuration properties of galaxies in the protocluster and the field, since we have applied an identical analysis to both. Nevertheless, we are now obtaining near-infrared *HST*-WFC3 observations that will allow us to repeat our analysis based on the morphology of the rest-frame optical flux. Similarly, we have ALMA follow-up observations of LBGs along the 'merger sequence' in SSA22 that will allow us to directly detect the thermal emission from cold dust responsible for extinction in the ultraviolet/optical light.

Our result is marginal mainly due to small number statistics, which dominate the size of the uncertainties on our merger fractions. This arises from the lack of a complete LBG catalogue covering the whole SSA22 field and sparse *HST* coverage of SSA22. It would be useful to extend our analysis to additional protoclusters and fields at a similar redshift to overcome the potential issues of small number statistics and cosmic variance, but high-$z$ protoclusters are extremely rare, and generally lack the ancillary data that SSA22 has. Moreover, it is important to note that the morphological classification approach we have presented here *must* be applied to identically selected galaxy populations in order to control for the influence of galaxy properties (chiefly mass) when investigating the influence of large-scale environmental effects in this way. Again, this is not possible for the majority of known $z > 2$ protoclusters.

Currently, the only other LBG catalogue in Steidel et al. (2003) with *HST* F814W coverage is the Westphal field. The *HST* data available here are generally shallower than those available for SSA22 and HDF-N and we have therefore not included them in our main analysis. However, we have classified the LBGs in this field where possible, following the same procedure as described above. We measure a merger fraction of $30 \pm 5$ per cent (26 mergers out of 88 LBGs for which images of reasonable quality were available). This is consistent with our main results for the HDF-N and SSA22 field samples and, if included, would result in a combined average field merger fraction of $30 \pm 3$ per cent. In order to solidify our result further we would need to significantly improve the spectroscopic completion of the protocluster membership to at least $\sim 70$ (LBG) members for a $3\sigma$ measurement.

## 5 SUMMARY

We have detected a marginal enhancement of the merger fraction of LBGs in the SSA22 protocluster at $z = 3.1$, with approximately 60 per cent more LBGs actively undergoing mergers in the protocluster compared to the average density field at the same epoch, consistent with large-scale simulations. This hints that mergers could have an important role in driving accelerated galaxy growth in the rare, dense environments destined to become the massive clusters of the present epoch, by triggering and driving star formation and black hole growth.

## ACKNOWLEDGEMENTS

We thank the anonymous referee for comments that have improved this paper. We thank Chris Conselice and Ian Smail for useful discussions and comments and Neil Cook and Emma Lofthouse for helpful assistance. NH is supported by the Science and Technology Facilities Council. JEG. is supported by a Royal Society University Research Fellowship. YM acknowledges support from Japan Society for the Promotion of Science KAKENHI grant number 20647268. All of the data presented in this paper are publicly available through the Mikulski Archive for Space Telescopes (MAST). STScI is operated by the Association of Universities for Research in Astronomy, Inc., under NASA contract NAS5-26555. This research made use of the following software packages: AS-TROPY (Astropy Collaboration et al. 2013); MATPLOTLIB (Hunter 2007); SCIPY (Jones & et al. 2001); MONTAGE (NASA/IPAC); The KAPETYN Package (Terlouw & Vogelaar 2012); TOPCAT (Taylor 2005); IRAF (Tody 1993); COSMOLOPY.

This paper has been typeset from a TeX/LaTeX file prepared by the author.